\documentclass[12pt]{article}

\usepackage{graphicx}

\title{Understanding two slopes in the $ pp (p\bar p)$ differential cross sections}

\author{    Yu.A.Simonov \\
 NRC ``Kurchatov Institute'' -- ITEP,\\ B. Cheremushkinskaya 25, \\Moscow, 117259, Russia}

\newcommand{\beq}{\begin{eqnarray}}
 \newcommand{\eeq}{\end{eqnarray}}
 \newcommand{\be}{\begin{equation}}

 \newcommand{\ee}{\end{equation}}

 \def\la{\mathrel{\mathpalette\fun <}}
\def\ga{\mathrel{\mathpalette\fun >}}
\def\fun#1#2{\lower3.6pt\vbox{\baselineskip0pt\lineskip.9pt
\ialign{$\mathsurround=0pt#1\hfil ##\hfil$\crcr#2\crcr\sim\crcr}}}

\newcommand{{\SD}}{\rm SD}

\newcommand{{\Mc}}{\mathcal{M}}

\newcommand{\ver}{\mbox{\boldmath${\rm r}$}}

\newcommand{\vep}{\mbox{\boldmath${\rm p}$}}
\newcommand{\veq}{\mbox{\boldmath${\rm q}$}}
\newcommand{\veQ}{\mbox{\boldmath${\rm Q}$}}

\newcommand{\velambda}{\mbox{\boldmath${\rm \lambda}$}}
\newcommand{\veR}{\mbox{\boldmath${\rm R}$}}

\newcommand{\vek}{\mbox{\boldmath${\rm k}$}}

\newcommand{\vexi}{\mbox{\boldmath${\rm \xi}$}}
\newcommand{\veta}{\mbox{\boldmath${\rm \eta}$}}

\begin{document}
\maketitle

\begin{abstract}
Recent experiments have discovered two exponents in the  $pp$ elastic differential cross sections with two different slope parameters,
of the order $(16-20)$~GeV$^{-2}$  and $(4-4.8)$~GeV$^{-2}$ in the regions $  -t \la 0.5 $~GeV$^2$ and
$ -t \ga 1$~GeV$^2$, respectively. We suggest a simple model of the $pp$ elastic scattering  with two types of particle
exchanges: 1) when the exchanged particle transfers the momentum $\veQ$ from a quark of the proton $p_1$ to one quark in another proton $p_2$, producing the slope $B_1$;
2) when the transfer occurs from two quarks in the $p_1$ to two quarks in the $p_2$, giving the exponent with the slope $B_2$. The resulting amplitude is proportional to
the product of the form factors of two protons, depending on $\veQ$, but with different coefficients  in the cases 1) and 2). Using  the only parameter - the proton
charge radius $r^2_{ch}= 0.93$~fm$^2$, one obtains $B_1 = 16$~GeV$^{-2}$,~$ B_2 = 4$~GeV$^{-2}$ with the strict value of the ratio,
$\frac{B_1}{B_2} = 4.0$, independent of $r_{ch}$. These predictions are surprisingly close to the data both in the $pp$ and in the $\bar p p$ differential cross sections. Comparison
to experimental data and theoretical approaches is discussed, together with possible implications for the future development of the theory.
\end{abstract}

\section{Introduction}

The important recent experiments of the TOTEM Collaboration on the $pp$ elastic scattering at high energies,  $E= 2.76$ TeV \cite{1}, $E = 7$ TeV \cite{2}, $E = 8$ TeV \cite{3},
and finally at $E = 13$ TeV \cite{4}, together with the proton-antiproton elastic differential cross sections, measured by the CDF Collaboration \cite{5}
at $E = 546$ GeV, by the E710 Collaboration at $E= 1.8$ TeV \cite{6}, by the D0 Collaboration at $E= 1.96$ TeV \cite{7}, provide very interesting material for theoretical
studies of possible mechanisms of the proton-proton and proton-antiproton interactions at high energy. The emergence of two very different exponents with a dip and
maximum between them calls for a formulation of a typical particle exchange picture with different types of exchanges in the first and the second regions. Moreover,
both slopes are almost purely exponential and their parameters are close to each other both in the $pp$ and the $p\bar p $ cases (around
16 GeV$^{-2}$ for the first slope) and slightly increase with the energy. Another enigma is the (almost) exact ratio $4:1$ for two slope parameters,
which was not yet explained by the existing mechanisms.

The idea of two exponentials with a relative phase for the scattering amplitude is not new and was used in many approaches, see e.g., \cite{8} and also \cite{9,10}.
One of the immediate questions is what is the nature of the second slope ? The answer to this question can be found in the known approaches, e.g. \cite{11,12,13},
in some substructure \cite{14}, or in the layered structure \cite{15,16,17}; specifically in \cite{17} the ratio of two slopes can be used to predict the sizes of two layers.
One of the possible scenarios considers two slopes as due to the quark-diquark structure of the proton \cite{12}. We shall come back to this topic in the discussion section of this paper.

\section{The model}

We consider the elastic proton-proton or antiproton-proton  scattering in the c.m. system with momenta
\be
P_1 (\vep) + P_2 (-\vep) \to P_3(\vep') + P_4(-\vep'),~~ \veQ = \vep' - \vep .
\label{eq.1}
\ee
We assume that the scattering proceeds via one-boson exchange mechanism (OBE) and the exchanged boson can transfer the momentum $\veQ$ from one proton to another via the
coupling (exchange) between the quarks in one proton and in another in two different ways: we call as the type 1), when the boson couples to only one quark in each proton,
and the type 2), when the boson couples to two quarks in each proton, or more precisely, it couples to the center of motion of two quarks,
e.g., $\frac{\ver_1 + \ver_2}{2}$. In this way the total amplitude can be written in a familiar way \cite{8},

\be
M = G_1(s,t) + G_2(s,t) = g_1(s,t) F_1(t) + g_2(s,t) F_2(t),~~ t= -\veQ^2,
\label{eq.2}
\ee
where $g_i(s,t)$ may contain an OBE pole, or a branch point, and multiple s-dependent corrections, while $F_i(t)$
is an OBE amplitude with two form factor vertices, containing the momentum transfer to the proton in the vertices (1,3) and (2,4),

\be
 f_{13}(Q): P_1(\vep) \to P_3(\vep') + b(\veQ);~~ f_{24}(Q): P_2(-\vep) + b(\veQ) \to P_4(-\vep').
\label{eq.3}
\ee
Both $f_{13}, f_{24}$ are equal to the proton form factor $f_i(Q)$ (here i refers to the slope $i=1,2$), which can be written via the proton wave functions as
\be
f_i(Q) = \sum_k \kappa_k \int d^3 \vexi d^3 \veta \psi(\vexi,\veta) \exp(i\veQ \velambda^{i}_k) \psi(\vexi,\veta).
\label{eq.4}
\ee
Here $\velambda^{(i)}_k$ denotes the coordinates, which accept or transfer the momentum $Q$: one quark coordinate
$\velambda^{(1)}_k = \ver_k$ for the first slope $(i=1)$ and the two-quark coordinate $\velambda^{(2)}= \frac{\ver_i + \ver_k}{2}$ for the second slope.
The coordinates $\vexi,\veta$ are introduced here in the same way as the internal coordinates in the
hyperspherical basis \cite{18}; in addition to the c.m. coordinate

\be
\veR = \frac{\ver_1 + \ver_2 + \ver_3}{3}, ~~\vexi= \frac{\ver_1 + \ver_2 - 2 \ver_3}{\sqrt{6}}, ~~\veta= \frac{\ver_2 - \ver_1}{\sqrt{2}}.
\label{eq.5}  \ee
An important property of the coordinates $\vexi, \veta$ is that the combination $\rho^2= \vexi^2 + \veta^2=
\frac{(\ver_1 - \ver_2)^2 + (\ver_1-\ver_3)^2 + (\ver_2 - \ver_3)^2}{3}$ does not contain any internal angular
momenta of the quarks, which can influence the $Q$ dependence of the form factors. It is clear that the ground states of the $3q$ system
have in the dominant component the wave function $\psi(\rho)$, which is called the hypercentral approximation in the framework of the hyperspherical approach,
developed in \cite{18}, originally in nuclear physics, and was used in the relastivistic formalism with the relativistic Hamiltonian in \cite{19}
to calculate the baryon masses \cite{19}, the magnetic moments \cite{20}, and the form factors \cite{21} in good agreement with experimental data.

We also define the corresponding momentum variables: $\veq= \frac{\partial}{i\partial \vexi},~ \vek= \frac{\partial}{i\partial \veta}$,
and also the proton wave functions in the momentum space in the simple Gaussian form, which is close to the wave function, calculated in \cite{19,20,21}
for protons and neutrons:

\be
 \psi(\vexi,\veta) = N \exp\left(-\frac{\mu^2(\vexi^2 + \veta^2)}{2}\right),~~\phi(\veq,\vek) = N'\exp\left(-\frac{\veq^2 +\vek^2}{2\mu^2}\right).
\label{eq.6}
\ee
These functions contain the only parameter $\mu$, which can be found from the proton charge or the magnetic radius,
as will be done below. Thus to find finally the proton form factors $f_i(Q)$ one needs to define in (\ref{eq.4})
the vector $\velambda^{(i)}_k$ in two situations: $i=1,i=2$. This is done as follows
\be
(1) i=1, \velambda^{(1)}_1= \ver_1,~~ \ver_1= \veR + \alpha_1 \vexi + \beta_1 \veta,~~ \alpha_1= \frac{1}{\sqrt6},~~
\beta_1= - \frac{1}{\sqrt2},
\label{eq.7}
\ee

\be
(2) i= 2, \velambda^{(2)}_1 = \frac{\ver_1 + \ver_2}{2} = \veR + \alpha_2 \vexi + \beta_2 \veta,~~ \alpha_2 = \frac{1}{\sqrt6},~~ \beta_2= 0,
\label{eq.8}
\ee
As a result the integral in (\ref{eq.4}) is calculated to be
\be
f_i(Q) = \exp\left( - \frac{Q^2 (\alpha^2_i + \beta^2_i)}{4\mu^2}\right),~~ \alpha^2_1 + \beta^2_1 = \frac{2}{3},~~
\alpha^2_2 + \beta^2_2 = \frac{1}{6}.
\label{eq.9}
\ee
In this way one immediately obtains the ratio of two exponential slopes in the total amplitude (\ref{eq.2})
\be
G_1(t) = f_1^2(t) = \exp\left(\frac{B_1 t}{2}\right),~~ G_2(t) = f_2^2(t) = \exp\left(\frac{B_2 t}{2}\right), ~~B_2 : B_1 = 1 : 4,
\label{eq.10}
\ee
which defines the slopes of the total differential cross section,
\be
\frac{d\sigma}{d t} = |g_1|^2 \exp(B_1 t) + |g_2|^2 \exp(B_2 t) + 2 {\rm Re}(f_1 f_2^*) \exp\left(\frac{(B_1+B_2)t}{2}\right).
\label{eq.11}
\ee
It is of interest to find the absolute values of $B_1,B_2$. To this end we define the proton electric (magnetic)
radius via $\mu$,
\be
r_p^2= - \frac{6}{f_1(0)} \frac{d f_1(Q)}{d Q^2}|_{Q=0} (hc)^2,~~ hc = 0.197~{\rm GeV~ fm},
\label{eq.12}
\ee
which yields
\be
\mu^2= \frac{(hc)^2}{r_p^2},~~ B_1= \frac{4}{6\mu^2} = \frac{2r_p^2}{3(hc)^2} = 17.18 r_p^2~ ({\rm GeV}^2)
\label{eq.13}
\ee
where the radius $r_p$ is in fm. Correspondingly, one obtains $B_1= 16$GeV$^{-2}$ for $r_p^2= 0.93$~fm$^2$, which is
$\sim 6\%$ larger than the standard value $(0.877\pm 13)$~fm$^2$ \cite{22}. One can already see that our results,

\be
B_1= 16 ~{\rm GeV}^{-2},~ B_2= 4 ~{\rm GeV}^{-2},
\label{eq.14}
\ee
are in the correct ballpark if one compares it with the experimental data \cite{1,2,3,4,5,6}.

\section {Comparison to experimental and theoretical data}

The main results of the previous section are the realistic derivation  of the two slopes $B_1,~B_2$ in the differential $pp$ and $p\bar p$ cross sections from the known value of the proton charge radius $r_p$, namely, $B_1= 16$~ GeV$^{-2},~ B_2= 4$~GeV$^{-2}$, which in our simple model do not depend on energy $E= \sqrt{s}$. In practice, however, both slopes can acquire
rescattering corrections and weakly depend on $E$. This indeed happens if one compares our numbers with
the experimental values of $B_1,B_2$ \cite{1,2,3,4}, given in Table~\ref{tab.01}, where
together with the values $B_1$ we give the approximate values of the slope $B_2$, whenever they can be estimated.

\begin{table}[!htb]
\caption{ The experimental values of slopes $B_1,B_2$  from the experiments \cite{1,2,3,4}}
\begin{center}
\label{tab.01}
\begin{tabular}{|l|c|c|c|c|}
\hline
$E $ (in TeV)       & 2.76       & 7           & 8          & 13 \\

References        & \cite{1}     & \cite{2}   & \cite{3}    & \cite{4} \\
\hline

$B_1$~(in GeV$^{-2}$) & 17.1-19.4    & ~21      &$ 19.35 \pm 0.06$ &$ 20.40 \pm 0.3$ \\

$B_2$ (in GeV$^{-2}$) & $\sim 4.45$  &$\sim 4.6$  &               & 4.6   \\

\hline
\end{tabular}
\end{center}
\end{table}

It is interesting to compare these values of $B_1,~B_2$ with the corresponding values from the $p\bar p$ differential
cross sections at smaller energies. For example, in the CDF data \cite{5} at $E= 0.546$ TeV one obtains $B_1 = (16.98 \pm 0.25)$ GeV$^{-2}$,
while from the E710 data \cite{6} one gets $B_1 = (16.1 \pm 0.3 )$~GeV$^{-2}$ and the similar value from the D0
experiment: $B_1= (16.8 \pm 0.4)$~GeV$^{-2}$, which are in agreement with our simple model prediction.
Summarizing one can see a qualitative and a rough quantitative agreement of both $pp$ and $p\bar p$ data with
our model. In addition one can also see a growing  with $E$ the first slope $B_1$, changing  by $(10-15)\%$ percent from
$E= 2.76$~TeV to $E= 13$~TeV. This increase can be estimated in the pomeron-exchange picture \cite{23}, where $\Delta B_1= 4 \alpha'_P(0) log(\frac{s}{s_1})$,
which yields \cite{23} the  $B_1$ changing from 16.8 GeV$^{-2}$ to 21.4 GeV$^{-2}$ on the interval $E= (1.8 - 13)$ TeV. It allows to understand the main features of the first slope behavior and
its energy dependence.

Another characteristic feature of the cross section is the position of the first dip which varies in different experiment \cite{1,4} from
0.6~GeV$^2$  to 0.47~ GeV$^2$  in the experiments with the energy interval (2.76-13) TeV.  To estimate the dip position one should compare the absolute magnitude of
two exponential terms with the coefficients, producing both slopes. One can assume that these coefficients are proportional to the squared matrix elements of the
wave function of the glueball (pomeron), interacting with one (the case $B_1$) or two quarks (the case $B_2$) inside proton. Then from (\ref{eq.1}) one can write the ratio

\be
\left|\frac{g_2}{g_1}\right|^2 = \left( \frac{r_{gl}}{r_p}\right)^n.
\label{eq.15}
\ee
Here $r_{gl}$ is the two-gluon glueball radius, which is in the range of 0.2~fm \cite{24} and therefore the value of the degree $n$ is equal to 6.0  from two  vertex integrations.
Therefore one can estimate $ - t_{\rm dip}= 2 \frac{{\rm log}\left(\frac{g_1}{g_2}\right)}{B_1 - B_2}$. As a result, one obtains a rough estimate $|t_{\rm dip}| = (0.5-0.8)$ GeV$^2$, which is in the correct ballpark,
indeed, in \cite{1} $|t_{\rm dip}|= 0.6$~GeV$^2$ and in \cite{4} $|t_{\rm dip}| = (0.47 \pm 0.01)$~GeV$^2$.

\section {Conclusions and an outlook}

The simple model suggested above is aimed to check the possible mechanisms beyond the surprising properties of
the differential $pp, p\bar p$ cross sections: namely, the almost invariant values of two slopes $(\sim 16,\sim 4)$~GeV$^{-2}$
and its ratio,  equal to 4.0, which were not presented in this combination by the models considered so far.
It was found in the paper that these specific numbers can be produced by the linear combination of two
exchanges, where in the first one the exchanged boson interacts with only one quark in both protons, while in the
second type of the exchange it interacts with two quarks in their center of the masses. This already provides the ratio of two slopes,
equal 4.0.

The explicit values of the slopes are obtained from the square of the product of two proton form factors
in the vertices of the OBE diagrams, which gives exactly 16~GeV$^2$ and 4~GeV$^{-2}$ for two slopes. One can consider this result as
an additional support of the quark-diquark \cite{12,14} and multilayer \cite{15,16} mechanisms of the
high energy $pp$ scattering.

It is remarkable that these results are in good agreement with experimental data in the wide region of energies
for both $pp$ and $p\bar p$ elastic scattering in the TeV region (however, differ in the GeV region).
The proposed exchanges can be provided by the two-gluon glueball exchanges in the pomeron series and therefore can be incorporated in the well developed
high energy scattering theory including the  reggeon formalism \cite{25,26}.

The author is grateful to A.M.Badalian for useful discussions. This work was supported by the Russian Science Foundation  in the framework of the scientific
project, grant 16-12-10414.

\end{document}